\documentclass[epj]{svjour}
\usepackage{latexsym}
\usepackage{graphicx}   
\begin{document}
\title{Soliton lattices originating from excitons interacting with high-intensity fields in finite molecular crystals}
\author{E. Nji Nde Aboringong \and Alain M. Dikand\'e\thanks{Email address: dikande.alain@ubuea.cm}
}                     
\authorrunning{Nji Nde Aboringong et al.}
\titlerunning{Exciton-polariton soliton trains in Kerr optical molecular crystals}

%
\institute{Laboratory of Research on Advanced Materials and Nonlinear Sciences (LaRAMaNS), Department of Physics, Faculty of Science, University of Buea, P.O. Box 63 Buea, Cameroon}
\date{Received: date / Revised version: date}
%
\abstract{
The effects of long-range intermolecular interactions on characteristic features of soliton bound states, consisting of localized excitons and polaritons in molecular crystals interacting with a high-intensity optical field, are investigated. Analytical solutions to the resulting modified nonlinear Schr\"odinger equation are obtained in terms of elliptic-type bright- and dark-soliton structures, which are assumed to correspond to periodic trains of pulse solitons and kink solitons respectively. Long-range intermolecular interactions are shown to renormalize the exciton-polariton interaction strength, hence generating a significantly huge increase in amplitudes of the bright solitons but a decrease in amplitudes of dark solitons. Results suggest that long-range intermolecular interactions hold relevant roles, both qualitatively and quantitatively, in the formation of amplitude and phase modulated strongly nonlinear exciton-polariton solitary-wave patterns, as well as in the energy transfer along molecular crystals interacting with high-intensity optical fields.
%
} 
\maketitle
\section{Introduction}
\label{intro}
\smallskip
The formation and propagation of solitary-wave structures in molecular systems have been widely investigated in the past four decades. Initial studies began with Davydov and Kislukha \cite{D73,Wei74} who showed that bound states of excitons, coupled with their self-induced lattice deformation in one-dimensional molecular chains, could propagate as solitary waves with constant shape and velocity. The latter, usually referred to as Davydov solitons, have been found useful for understanding the mechanisms by which energy is localized and transported in biopolymers and protein chains \cite{vp1,vp2,scott,Zhu,Xiao95,Xiao97,minga,cruz,tak1999,dan95,ex1}. In the previously cited works however, the generation and propagation of Davydov-type solitons were treated in the idealized picture where only first nearest-neighbour interactions between molecules are considered. Nevertheless, improvement to this idealized picture has been proposed relatively recently by taking into account the natural contribution of long-range intermolecular interactions \cite{grec,braun,woaf,vaz,alfi,yu96,yu97} on soliton characteristic properties, as well as their implications on energy and charge transports in biomolecular systems such as $\alpha$--helical proteins \cite{mvogLRI2013,mvog2013,edison1,tabi18}.
\smallskip
\newline
Bound states consisting of coupled exciton-photon solitons propagating in low-dimensional molecular lattices have also been investigated in the past, in connexion with a phenomenon known as self-induced transparency \cite{SI1,SI2,SI3,SI4,SI5,SI6,SI7,SI8,SI9,ex2}. In these systems, the formation of polariton solitons originates from exciton-exciton interactions balancing the polariton dispersion due to exciton-photon couplings. From a general standpoint, the formation of solitons in one-dimensional molecular crystals with nearest-neighbour intermolecular interactions, when excitons interact with low-intensity \cite{prim2001} and high-intensity \cite{prim2002} optical fields has been studied. The existence of both bright and dark solitons associated with large-amplitude excitons and polaritons in the system has been established \cite{prim2001,prim2002}. These bright and dark solitons are more precisely, localized wave structures forming in a molecular chain of an infinite length. Still for most real molecular systems and particularly biomolecular chains, the consideration of an infinite length turns out to be a shortcoming especially when considering applications in molecular systems with well-defined sizes. 
\smallskip
\newline
In a previous study \cite{edison2}, we examined the influence of long-range intermolecular interactions on bound exciton-polariton periodic soliton trains in one-dimensional molecular crystals in the case when the excitons were coupled to low-intensity optical fields. Worth noting, intermolecular forces in molecular crystals are usually hydrogen-bond interactions, van der Waals interactions and London-type forces and are mainly of a dispersive nature, being relatively weaker compared with typical intramolecular forces such as covalent and ionic bonds \cite{chem}. On the other hand, in some physical contexts the optical field interacting with excitons can be of relatively high intensity due to multi-photon correlations, and because of the related nonlinearity the dispersion of the optical field will be balanced which enhances the exciton-photon interactions leading to renormalization of the exciton-photon coupling strength. Therefore we expect a stronger lattice-field interactions, and consequently a stronger net nonlinearity in the system which will affect drastically characteristic features of periodic soliton trains associated both to excitons and polaritons.
\smallskip
\newline
In this paper, we examine the effects of long-range intermolecular interactions on characteristic parameters of exciton and polariton soliton trains in a finite-length molecular crystal interacting with a high-intensity field. In the next section we present the model, derive the equations of motion for both the excitons and polaritons and examine the effects of long-range intermolecular interactions on dispersion properties of the finite one-dimensional (1D) molecular chain. Here we consider a specific form of long-range intermolecular interaction potential which is peculiar in that it allows the number of interacting molecules to be finite, and eventuallyinferior to the total number of molecules in the chain. We obtain a modified nonlinear Schr\"odinger equation for the exciton wave function, in which the self-phase modulation nonlinear coefficient is renormalized by the long-range intermolecular interaction. In the subsequent section we solve the modified nonlinear Schr\"odinger equation assuming periodic boundary counditions consistent with the finiteness of the chain, and find period soliton-train solutions of the bright and dark soliton types. The work ends with a summary of results and concluding remarks.
\section{Model Hamiltonian and equations of motion}
\smallskip
Consider a linear molecular chain in which Frenkel excitons have formed and interact with a propagating electromagnetic field. The total Hamiltonian for such a system can be expressed \cite{prim2001,prim2002,edison2}:
\begin{eqnarray}
 H &=& \hbar{w}_0\sum_{n=1}^{N}{b}_{n}^{\dag}{b}_{n} -
 \sum_{n=1}^{N}\sum_{m \neq n}J_{n-m}{({b}_{n}^{\dag}{b}_{m} + {b}_{n}{b}_{m}^{\dag})} \nonumber \\
&-&
 D\sum_{n}^{N}{b}_{n}^{\dag}{b}_{n}{b}_{n+1}^{\dag}{b}_{n+1} -
 d\sum_{n=1}^{N}({b}_{n}^{\dag}e_{n}^{+} + b_{n}e_{n}^{-}), \label{hamLG}
\end{eqnarray}
where $\hbar{w}_0$ is the intramolecular excitation energy,
$J_{n-m}$ is the energy transfer integral between the molecular excitations
on sites $n$ and $m$. $D$ is the exciton-exciton coupling strength which would be studied in the
short range limit, assuming that interactions between exciton dimers along the chain are of short-range although single-exciton states can interact beyong the first nearest neighbours. The Pauli
operator ${b}_{n}^{\dag}$ $({b}_{n})$ creates (annihilates) a molecular
exciton at site $n$. The quantities $e^+$ and $e^{-}$ in equation (\ref{hamLG})
are the right-going and left-going components of the electromagnetic field,
where $d$ is the exciton-field coupling in the dipole approximation.

When $m=n\pm1$, the Hamiltonian in equation (\ref{hamLG}) reduces to that for a chain of excitons interacting with light where only nearest-neighbour interactions are considered \cite{prim2001,prim2002}. In the present study we consider the possibility for long-range intermolecular interactions between molecular units, by assuming a long-range interaction whose coupling strength falls off with a power law as one moves away from the first-neighbour molecules. One of such long-range interactions is the Kac-Baker potential \cite{dik96,kac63,bac63,dik95,dik07}:
\begin{equation}\label{kbp}
  J_{n-m} = {J}_0\frac{1-r}{2r}{r}^{|\ell|}, \qquad  \ell = m-n,
\end{equation}
where the parameter $r$  $(0 < r \leq1)$ determines the strength of
intermolecular interactions beyond the first neighbours, and $\ell$ is the
distance between a molecule at site $n$ and a molecule at site $m$. For the long-range interaction potential given in equation (\ref{kbp}), the first neighbour model \cite{prim2001,prim2002} is recovered when $r\rightarrow0$. When $\ell\rightarrow\infty$ (or $r\rightarrow1$), the intermolecular coupling becomes fully long-ranged. In this limit, dispersive intermolecular interactions are purely of van der Waals type  \cite{sark81}.

By its definition, the long-range interaction potential equation (\ref{kbp}) applies only to physical contexts for which $\ell\rightarrow1$ or $\ell\rightarrow\infty$ \cite{kac63,bac63,sark81}. However, in most real physical contexts long-range interactions among molecules are effectively sensitive only over a finite range of intermolecular distances, saturating beyond. This means that $\ell$ should be finite and consequently the Kac-Baker potential becomes unappropriate. As an alternative we shall consider a modified version of this potential, which takes into consideration the finiteness of the interaction range i.e. \cite{dik07}:
\begin{equation}\label{mkbp}
   J_{n-m}^{L} = \frac{{J}_0}{1-r^L}\frac{1-r}{2r}{r}^{|\ell|}, \qquad  \ell = m-n,
\end{equation}
where L is the maximum range beyond which intermolecular interactions
saturate. In the nearest-neighbour limit i.e. $\ell=L=1$, the
modified Kac-Baker potential in equation (\ref{mkbp}) reduces to $J_1 = {J_0}/2$ irrespective of the
value of $r$. In the same nearest-neighbour limit, the Kac-Baker potential
becomes $J_1=J_0(1-r)/2$. So for the Kac-Baker potential, we recover the
nearest-neighbour value of the intermolecular interaction strength only when $r=0$.

\begin{figure}[h!]
\centering
\includegraphics[width=3.3in]{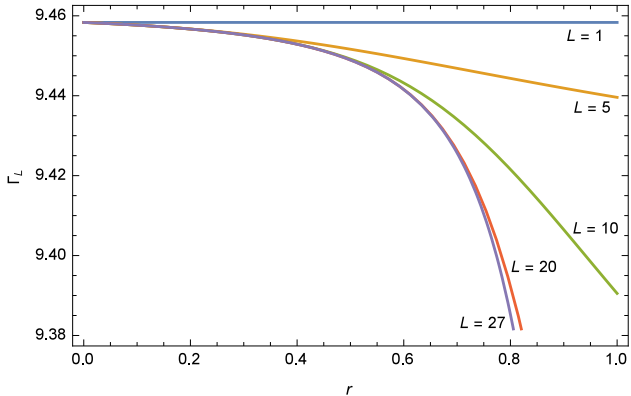}
  \caption{The nonlinear coefficient $\Gamma_L$ in the long-wavelength regime (q $\approx$ 0.05),
  plotted versus the long-range interaction strength $r$ for five different values of the interaction range i.e.
  $L = 1$, $5$, $10$, $20$ and $27$.
 }
  \label{fig:Fig1}
\end{figure}

\begin{figure}[h!]
\centering
\includegraphics[width=3.3in]{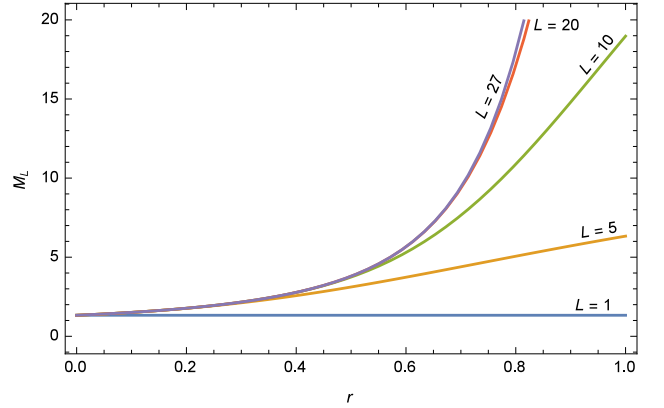}
  \caption{The dispersive coefficient $M_L$ in the long-wavelength regime (q $\approx$ 0.05),
  plotted versus the long-range interaction strength $r$ for five different values of the interaction range i.e.
  $L = 1$, $5$, $10$, $20$ and $27$.
 }
  \label{fig:Fig2}
\end{figure}

The Pauli operators ${b}_{n}^{\dag}$ and ${b}_{n}$ prohibit the localization of more than one
exciton on a single molecule and obey the commutation relations:
\begin{eqnarray}
  [{b}_{n},{b}_{m}^{\dag}] = (1 - 2P_n)\delta_{n,m}, \qquad [{b}_{n},{b}_{m}] = 0 \nonumber \\
  {b}_{n}^2 = ({b}_{n}^{\dag})^2 = 0, \qquad P_n \equiv {b}_{n}^{\dag}{b}_{n}. \label{pauli}
\end{eqnarray}
Applying the Heisenberg formalism on the Hamiltonian formula (\ref{hamLG}) with the help of the commutation relations in (\ref{pauli}), we obtain the exciton equation of motion:
\begin{eqnarray}
 i\hbar\frac{\partial b_n}{\partial t} &=& \hbar\omega_0{b_n} - (1 - 2P_n)\sum_{m \neq n}{J}_{n-m}^{L}b_m \nonumber \\
&-&
  2Db_n\left({b}_{n+1}^{\dag}{b}_{n+1} + {b}_{n-1}^{\dag}{b}_{n-1}\right) \nonumber \\
  &-& d(1 - 2P_n)e_{n}^{+}. \label{eomPL}
\end{eqnarray}
Due to our interest in coherent structures associated with the
dynamics of single-particle states, we shall focus on the expectations of the Pauli operators as well as of their products. In this respect we can set
$\langle P_n \rangle \approx {|\langle b_n \rangle|}^2$, and define $\langle
b_n\rangle = \alpha_n$ where $\alpha_n$ is the single-exciton occupation
probability \cite{dik07epj}. With theses equation (\ref{eomPL}) becomes:
\begin{eqnarray}
 i\hbar\frac{\partial \alpha_n}{\partial t} &=& \hbar\omega_0{\alpha_n} - \left(1 - 2|\alpha_n|^2\right)\sum_{m \neq n}{J}_{n-m}^{L}b_m \nonumber \\
&-& 2D\alpha_n\left(|\alpha_{n+1}|^2 + |\alpha_{n-1}|^2\right)\nonumber \\
&-& d\left(1 - 2|\alpha_n|^2\right)e_{n}^{+}. \label{eomPL1}
\end{eqnarray}
We postulate that the optical response of the medium to the propagating electromagnetic field $e_n$ is nonlinear for strong lattice-field interactions, such that the field propagation can be described by the nonlinear Maxwell equation with a cubic (i.e. Kerr) nonlinearity:
\begin{eqnarray}
\left(\frac{\partial^2}{\partial x^2} - \frac{1}{c^2}\frac{\partial^2}{\partial t^2} \right)e^{+}(x,t) &=&
4\pi \frac{\partial^2}{\partial t^2}[\frac{d}{a^3}\alpha(x,t) + \nonumber \\
&&\chi^{(3)}|e^{+}(x,t)|^2e^{+}(x,t)], \label{MaxnonLin}
\end{eqnarray}
where ``$a$" is the molecular lattice spacing, $c$ is the speed of light (which shall be set
to unity) and $\chi^{(3)}$ is the third order nonlinear optical
susceptibility. The first term on the right-hand side of equation (\ref{MaxnonLin}) accounts
for the light-induced linear polarization of the molecular lattice, while the
second term describes the light-induced Kerr-type nonlinear polarization of
the same medium. \\
The single-exciton occupation probability equation (\ref{eomPL1}) and the nonlinear Maxwell equation (\ref{MaxnonLin}), form a set of
equations which describe the properties of nonlinearly coupled excitons and
photons and this model has been recently studied \cite{prim2002}, where  only first-neighbour intermolecular interactions were considered. We are interested in the effects
which long but finite-range dispersive intermolecular interactions would have on the periodic trains of exciton and polariton
soliton excitations, when the optical response of the medium is nonlinear. Proceeding with let us seek solutions to these equations of the following forms:
\begin{equation}\label{s1}
\alpha(x,t) = e^{i(qx-\omega t)}\phi(x, t), \qquad e^{+}(x,t) = e^{i(qx-\omega t)}\varepsilon(x, t),
\end{equation}
where $\omega$ and $q$ represent the carrier wave frequency and wavenumber respectively. Treating the exciton equation of motion (with $\hbar\equiv 1$) in the continuum limit, and making use of the modified Kac-Baker potential in the same limit with the help of formula (\ref{s1}), equations (\ref{eomPL1}) and (\ref{MaxnonLin}) reduce to:
\begin{eqnarray}
   i\frac{\partial \phi(x,t)}{\partial t}   &=& (\epsilon_{L}(q) - \omega)\phi(x,t) - iv_{L}(q)(1 - 2|\phi|^2)\frac{\partial \phi(x,t)}{\partial x} \nonumber\\
       &-& b^{(a)}_{L}(q)\frac{\partial^2 \phi(x,t)}{\partial x^2}
       - 4(D - b_{L}(q))|\phi|^2\phi(x,t) \nonumber\\
       &-&\frac{i}{6}v^{(a)}_{L}(q)\frac{\partial^3
\phi(x,t)}{\partial x^3}
- d(1 - 2|\phi|^2)\varepsilon(x,t), \label{NLeq1}
\end{eqnarray}

\begin{eqnarray}
   [\left(\omega^2 - q^2\right) &+&
2i\left(q\frac{\partial}{\partial x} + \omega\frac{\partial}{\partial t}\right) + \left(\frac{\partial^2}{\partial x^2} -
\frac{\partial^2}{\partial t^2} \right) ]\varepsilon(x,t) \nonumber \\
      &=&4\pi\left(- \omega^2 - 2i\omega\frac{\partial}{\partial t} +
\frac{\partial^2}{\partial t^2}\right)[\frac{d}{a^3}\phi(x,t) \nonumber \\
&+& \chi^{(3)}|\varepsilon|^2\varepsilon(x,t)], \label{NLeq2}
\end{eqnarray}
with:
\begin{equation}\label{wql}
\epsilon_{L}(q) = \omega_0 - 2b_{L}(q),
\end{equation}
\begin{equation}\label{bql}
b_{L}(q) = \sum_{\ell=1}^{L}J_{\ell}\cos(\ell qa), \qquad
b^{(a)}_{L}(q) = -\frac{\partial^2 b_L(q)}{\partial q^2},
\end{equation}
\begin{equation}\label{vql}
v_{L}(q) = \frac{\partial\epsilon_L(q)}{\partial q}, \qquad
v^{(a)}_{L}(q) = -\frac{\partial^3\epsilon_L(q)}{\partial q^3}.
\end{equation}
The quantities $\epsilon_L(q)$ and $v_{L}(q)$ stand for the energy and velocity
of single-exciton states respectively. Replacing the expression for the modified Kac-Baker potential
(\ref{mkbp}) in that for $b_L(q)$, the sum over $\ell$ can be evaluated
exactly \cite{dik07}. So the expression for $\epsilon_L(q)$ yields exactly:
\begin{equation}\label{sumag1}
\epsilon_L(q) = \omega_0 - J_0\frac{1-r}{1-r^L}\frac{\cos(qa) - r - r^LK_L(q)}{1 - 2r\cos(qa) + r^2},
\end{equation}
\begin{equation}\label{sumag2}
K_L(q) = \cos\left[(L + 1)qa\right] - r \cos\left(qLa\right).
\end{equation}
The effects of long-range intermolecular interactions on the energy and velocity of single-exciton states have been well established and illustrated in \cite{edison2}. It is important to note that lattice-field interactions may not introduce any quantitative or qualitative change on the velocity and energy spectra of single-exciton states since the latter represent exciton states with zero photon interactions. In equation (\ref{NLeq1}) we have kept the nonlinear dispersion term
$\sim|\phi|^2\partial\phi/\partial x$ and the third-order linear dispersion
term $\sim\partial^3\phi/\partial x^3$. We introduce the new
coordinate $z = x - vt$, and apply the slowly varying envelope approximation \cite{prim2001}. Also expressing the electric field from (\ref{NLeq1}) as
a function of the polarization, using the small-amplitude approximation $(1-2|\phi|^2)^{-1} \simeq 1 + 2|\phi|^2$, and substituting this into equation (\ref{NLeq2}), keeping
first-order nonlinear dispersion and third-order linear dispersion terms, we
obtain a modified version of the nonlinear Schr\"odinger equation for the
polarization \cite{prim2002} given as:
\begin{eqnarray}
 P_L\phi - M_L\frac{\partial^2\phi}{\partial z^2} &-& \Gamma_L|\phi|^2\phi - i(S_L - T_L|\phi|^2)\frac{\partial \phi}{\partial z} \nonumber \\
 &+& iQ_L\phi^2\frac{\partial \phi^{*}}{\partial z} - iR_L\frac{\partial^3 \phi}{\partial z^3} = 0,\label{MNLS}
\end{eqnarray}
\begin{equation}\label{meom22}
 \frac{\partial^2\varepsilon}{\partial z^2} = \frac{1}{d}\left[(\epsilon_L(q) - \omega)\frac{\partial^2\phi}{\partial z^2}-i(v_L(q) - v)\frac{\partial^3\phi}{\partial z^3}\right], \label{fie1}
\end{equation}
where the coefficients are defined as follows:
\begin{eqnarray}
  P_L &=& \epsilon_L(q) - \omega - \frac{\Omega_0\omega^2}{q^2 - \omega^2}, \nonumber \\
  M_L &=& b^{(a)}_L(q) \nonumber \\
  &+& \frac{2(q-\omega v)(v_L(q) - v) + (1 - v^2)(\epsilon_L(q)-\omega) - \Omega_0v^2}{q^2 - \omega^2}, \nonumber \\
  \Gamma_L &=& 4D - 2(\omega_0 - \omega) + \chi\frac{\omega^2(\epsilon_L(q) - \omega)^3}{q^2 - \omega^2}, \nonumber \\
  S_L &=& (v_L(q) - v) + \frac{2[(q-\omega v)(\epsilon_L(q) -\omega) - \Omega_0\omega v]}{q^2 - \omega^2}, \nonumber \\
   T_L &=& 2v \nonumber \\
   &+& \frac{[8(2D - \omega_0 + \omega)(q - \omega v)]}{q^2 - \omega^2} \nonumber \\
&+& \frac{[2\chi\omega (\epsilon_L(q) - \omega)^2(2v(\epsilon _L(q) - \omega)-\omega(v_L(q) - v))]}{q^2 - \omega^2}, \nonumber \\
  Q_L &=& \frac{[4(2D -\omega_0 + \omega)(q - \omega v)]}{q^2 - \omega^2} \nonumber \\
  &+& \frac{[\chi\omega(\epsilon_L(q) -\omega)^2(2v(\epsilon_L(q)-\omega) -\omega(v_L(q) - v))]}{q^2 - \omega^2}, \nonumber \\
  R_L &=& \frac{v^{(a)}_L(q)}{6} - \frac{2(q-\omega v)b^{(a)}_L(q) + (1-v^2)(v_L(q) - v)}{q^2 - \omega^2}, \nonumber \\
  \Omega_0 &=& \frac{4\pi d^2}{a^3}, \qquad \chi = \frac{4\pi \chi^{(3)}}{d^2}. \label{TERMS}
\end{eqnarray}
Equations (\ref{MNLS}) and (\ref{fie1}) govern the
formation and propagation of amplitude modulated and phase modulated
exciton-polariton solitons in the one-dimensional molecular system with dispersive, but finite long-range intermolecular interactions, as reflected in the dependence of the coefficients (\ref{TERMS}) on characteristic long-range parameters $L$ and $r$. In order to illustrate this dependence, in figs. \ref{fig:Fig1} and \ref{fig:Fig2} we plot the effective nonlinear coefficient $\Gamma_L$ and the dispersion coefficient $M_L$ in the long-wavelength limit ($q \approx
0.05$), versus $r$ for different values of the interaction range $L$.

For figs. (\ref{fig:Fig1}) and (\ref{fig:Fig2}), we used the following arbitrary
parameter values; $J_0 = 1$, $\omega_0 = 0.9$, $\omega = 1.2$, $\Omega_0 =
0.1$, $a = 1$, $\chi = 0.3$ and $v = 0.2$. The horizontal curve in fig. (\ref{fig:Fig1}) corresponds
to the effective nonlinear coefficient in the short-range regime, where only the first nearest
neighbour interaction are considered ($L = 1$). The nonlinear coefficient $\Gamma_L$ tends
to decrease from its short range value, to a threshold value below
which it decreases no further as $L$ exceeds a finite critical value (of $27$ nearest neighbours in this particular illustrative case).
In fig. \ref{fig:Fig2}, where the dispersive coefficient
$M_L$ is plotted versus $r$ in the long-wavelength regime for five different values of the
interaction range $L$, one sees that as $r$ is increased the dispersive coefficient increases
from its short-range value. The dispersion fully saturates as $L$ exceeds the same critical value found for the effective nonlinear coefficient $\Gamma_L$.

\section{Solitary wave train solutions in the continuum}
\smallskip
We seek solitary-wave solutions to equation (\ref{MNLS}) with real amplitudes and chirps, given by:
\begin{equation}\label{ansz}
\phi(z) = \rho(z)\exp[{i\Phi(z)}],
\end{equation}
where $\rho(z)$ is the solitary wave amplitude and $\Phi(z)$ is the chirp representing a nonlinear
contribution to the phase. Separating real and imaginary parts of equation
(\ref{MNLS}), we obtain a system of coupled nonlinear equations:

\begin{eqnarray}
  {\rho}[P_L &+& M_L\left(\frac{\partial \Phi}{\partial z}\right)^2] - \left[M_L - 3R_L\frac{\partial \Phi}{\partial z}\right]\frac{\partial^2\rho}{\partial z^2} -
  {\rho^3}\Gamma_L  \nonumber \\
 &+& {\rho}[S_L - (T_L-Q_L)\rho^2]\frac{\partial \Phi}{\partial z}  + 3R_L\frac{\partial \rho}{\partial z}\frac{\partial^2 \Phi}{\partial z^2} \nonumber \\
  &+& \rho{R_L}\frac{\partial^3\Phi}{\partial z^3} = 0, \label{C1}
\end{eqnarray}

\begin{eqnarray}
{\rho}{M_L}\frac{\partial^2\Phi}{\partial z^2} &+& \left[S_L + 2M_L\frac{\partial \Phi}{\partial z}\right]\frac{\partial \rho}{\partial z} -
\rho^2(T_L + Q_L)\frac{\partial \rho}{\partial z} \nonumber \\
&+& R_L\frac{\partial^3 \rho}{\partial z^3} = 0. \label{C2}
\end{eqnarray}
To solve equations (\ref{C1}) and (\ref{C2}), we approximate the third-order derivative
$\partial^3\rho/\partial z^3$ by the following expression derived from the
cubic nonlinear Schr\"odinger equation:
\begin{equation}\label{r3}
\frac{\partial^3 \rho}{\partial z^3} = \frac{P_L - 3\rho^2\Gamma_L}{M_L}\frac{\partial \rho}{\partial z}.
\end{equation}
Replacing equation (\ref{r3}) in (\ref{C2}) and integrating twice, we obtain
an expression for the phase derivative:
\begin{equation}\label{phiprim}
  \frac{\partial \Phi}{\partial z} = -\frac{{S_L}M_L + {R_L}{P_L}}{2M_L^2} + \left[\frac{({T_L}+{Q_L})M_L + 3{R_L}{\Gamma_L}}{4M_L^2}\right]\rho^2.
\end{equation}
Integrating equation (\ref{r3}) once with respect to $z$, substituting the
results into (\ref{C1}) alongside the phase derivative satisfying (\ref{phiprim}) such that the constant term vanishes, we obtain the following cubic-quintic nonlinear Schr\"odinger equation for the exciton amplitude
$\rho$:
\begin{equation}\label{quint}
 {P_L}\rho - {M_L}\frac{\partial^2 \rho}{\partial z^2} - {G_L}\rho^3 - {F_L}\rho^5 = 0,
\end{equation}
with the coefficients defined as:
\begin{eqnarray}
{G_L} &=& \Gamma_L - \frac{{R_L}{P_L}}{2M_L^3}[({T_L}+{Q_L}){M_L} + 3{R_L}{\Gamma_L}], \nonumber \\
  {F_L} &=& \frac{1}{16M_L^3}[(3{T_L}-5{Q_L}){M_L} + 33{R_L}{\Gamma_L}][({T_L}+{Q_L}){M_L} \nonumber \\
   &+& 3{R_L}{\Gamma_L}]. \label{eqard}
\end{eqnarray}

\subsection{Bright Soliton trains}
With exception of long-range effects which were introduced in the present work, it is known \cite{prim2002} that when $M_L$ and $G_L$ are of the same sign equation (\ref{quint}) admits the single-pulse solution:
\begin{equation}\label{Nsol1}
\rho (z) = \rho_0\sqrt{\frac{1 + B_L}{1 + {B_L}\cosh (2z/\tilde{L})}}, 
\end{equation}
with the carrier frequency $\omega$ obeying the transcendental relation:
\begin{equation}\label{ND1}
\omega = \epsilon_L(q) - \frac{G_L\rho_0^2(1+B_L)}{4} - \frac{\Omega_0\omega^2}{q^2 - \omega^2},
\end{equation}
and where we defined:
\begin{equation}\label{Ltild}
  \tilde{L}^2 = \frac{4M_L}{G\rho_0^2(1+B_L)},
\end{equation}
and,
\begin{equation}\label{B}
  B_L = 1 + \frac{4F_L}{3G_L}\rho_0^2>0.
\end{equation}
The quantity $\tilde{L}$ is the pulse width, and $B_L$ is a pulse characteristic parameter that depends on the arbitrary integration constant $\rho_0$ in addition to the characteristic long-range parameters $L$ and $r$. If for some values of characteristic parameters of the model $F_L$ can go to zero, then $B_L=1$ and the solution (\ref{ND1}) coincides with solution to the standard cubic nonlinear Schr\"odinger equation with an amplitude $\rho_0$ \cite{scott}.

Most generally, equation (\ref{quint}) is a fifth-order elliptic ordinary differential equation. Therefore it can admit elliptic-function solutions which, for
the bright-soliton condition (i.e .$M_L$ and $G_L$), is a pulse soliton train given by:
\begin{eqnarray}\label{PeA}
  \rho_\nu (z) &=& \rho_0\sqrt{\frac{1+B_L}{1 + 2{B_L}\kappa^2cn^2\left[dn^{-1}\left(\cos(iz/\tilde{L})\right)\right]}}, \nonumber \\
  B_L &>& 0.
\end{eqnarray}
In the latter solution $cn()$ and $dn()$ are Jacobi elliptic functions of modulus $\kappa$ \cite{dikand,dikand1,dikand2},
lying within the interval $0<\kappa\leq1$. It is remarkable that the
soliton feature of the solution (\ref{PeA}) gets lost as $\kappa \rightarrow
0$ \cite{dik96}, a limit where the Jacobi elliptic functions tend to sinusoidal functions. On the other hand, when $\kappa \rightarrow 1$ the pulse-train solution (\ref{PeA}) is transformed exactly to the single-pulse soliton obtained in (\ref{Nsol1}).

\subsection{Dark Soliton trains}
When the parameters $M_L$ and $G_L$ are of different signs, equation
(\ref{quint}) admits a dark soliton solution which was proposed in ref. \cite{prim2002} as being:
\begin{equation}\label{Nsol2}
\rho(z) = \frac{\sqrt{\rho_1B_L}}{\sqrt{cosech^2(z/\tilde{L}) +B_L}},
\end{equation}
with the carrier frequency $\omega$ obeying the transcendental relation:
\begin{equation}\label{ND2}
\omega = \epsilon_L(q) - \frac{G_L(1-3B_L)\rho_1^2}{2(2-3B_L)} - \frac{\Omega_0\omega^2}{q^2 - \omega^2},
\end{equation}
and where we defined:
\begin{equation}\label{Ltild}
 \tilde{L}^2 = \frac{2{M_L}(2-3B_L)}{G_L\rho_1^2}.
\end{equation}

The elliptic-function solution to equation (\ref{quint}), in the
dark-soliton regime, is obtained by integrating this fifth-order elliptic ordinary differential equation with appropriate sign constraints on the effective nonlinear coefficient and the dispersive coefficient. This yields: 
\begin{equation}\label{PeB}
\rho (z) = \rho_1\sqrt{\frac{B_L}{B_L - 2ds^2\left[cn^{-1}\left(\frac{\kappa^\prime}{\kappa}\cot(-i\kappa \kappa^\prime z/2\tilde{L})\right)\right]}}, 
\end{equation}
with:
\begin{equation}\label{kappa}
  \kappa^\prime = \sqrt{1-\kappa^2}.
\end{equation}

The function $ds()$, which represents the ratio of $dn()$ and $sn()$ functions \cite{abra}, is
another Jacobi elliptic function of modulus $\kappa$ lying within the
interval $0<\kappa<1$. As $\kappa
\rightarrow 1$, the dark elliptic-function solution coincides exactly with the single dark soliton structure (\ref{Nsol2}).

\bigskip

\begin{figure}
\centering
\includegraphics[width=\linewidth]{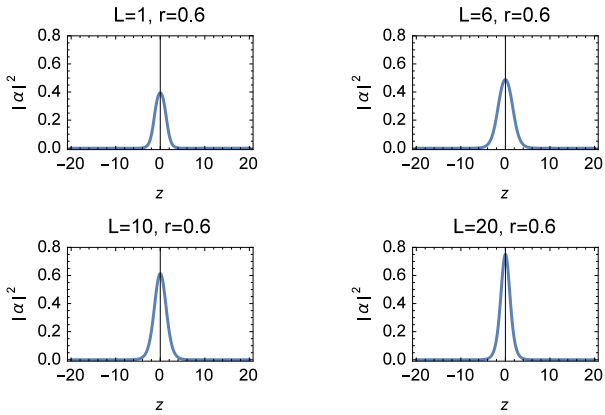}
  \caption{ The squared modulus of the bright soliton solution versus $z$ for four different values of the interaction range, $L$, i.e. $L=1$, $6$, $10$, $20$: $r=0.6$.
 }
  \label{fig:Fig3}
\end{figure}

\begin{figure}
\centering
\includegraphics[width=\linewidth]{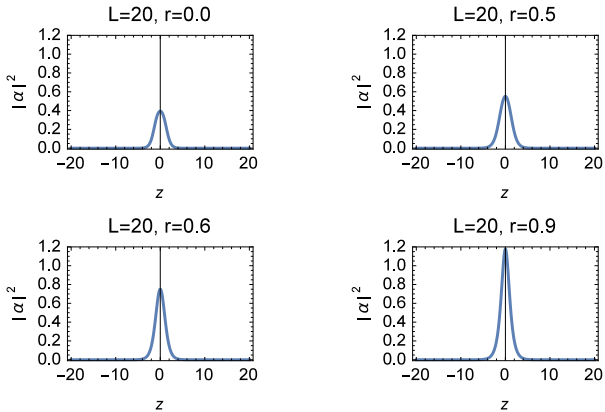}
  \caption{ The squared modulus of the bright soliton solution versus $z$ for four different values of the interaction strength, $r$, i.e. $r=0.0$, $0.5$, $0.6$, $0.9$: $L=20$.
 }
  \label{fig:Fig4}
\end{figure}

In fig. \ref{fig:Fig3}, the squared modulus of the occupation probability $\alpha$ for the exciton with a bright soliton amplitude (\ref{Nsol1}), is
sketched versus $z$ for four different values of the intermolecular interaction range i.e. $L = 1$, $6$, $10$ and $20$. Here the interaction strength is fixed to $r=0.6$. In fig. \ref{fig:Fig4} we also plotted the squared modulus of the occupation probability for a bright-amplitude exciton, when $L = 20$ and considering four different values of the interaction strength i.e. $r=0.0$, $0.5$, $0.6$, $0.9$. Our aim in showing these two figures is to bring out the distinct changes experienced by the exciton amplitude when the interaction strength $r$ is varied for a fixed intermolecular interaction range $L$ on one hand, and when the interaction range is varied while the interaction strength is held fix on the other hand. It is equally worthwhile stressing that the elliptic-function solutions obtained in the bright and dark soliton conditions, are periodic trains of the single-bright and single-dark soliton solutions (\ref{Nsol1}) and (\ref{Nsol2}) respectively. So for the graphical analysis we find it more enriching focusing on one individual component of each of the two soliton-train solutions. Indeed this will enable us gain insight onto the changes caused by variations of the two characteristic long-range interaction parameters $L$ and $r$, on the exciton amplitudes. \\

For  fig. \ref{fig:Fig3} and fig. \ref{fig:Fig4} we used the following arbitrary values for the model parameters: $J_0 = 1$, $\omega_0 = 0.9$, $\omega = 1.2$, $\Omega_0 =
0.1$, $a = 1$, $\chi = 2.0$, and $v = 0.2$. Instructively the case $L = 1$ corresponds to the first-neighbour model, in which the bright soliton solution maintains a steady pulse shape irrespective of the value of the intermolecular coupling strength $r$. This is the case that was studied in ref. \cite{prim2002}. As the intermolecular interaction range $L$ is gradually increased from the first nearest neighbours (fig. \ref{fig:Fig3}), it is observed that the bright soliton profile undergoes a significantly huge increase in amplitude. As the strength of intermolecular interactions $r$ is gradually increased for $L=20$ (fig. \ref{fig:Fig4}), the bright soliton amplitude increases slightly in the region $0\leq r \leq0.5$ (i.e. for weak intermolecular coupling), and undergoes a significantly huge increase in the region $0.5< r\leq1$ (i.e. for strong intermolecular coupling). The drastic increase in the exciton amplitude with increasing values of the two characteristic long-range parameters, and particularly the quasi doubling of the exciton amplitude at relatively strong but finite long-range interations, is quite telltale of the significant role long-range intermolecular interactions are expected to play in the quantitative estimate of the energy effectively conveyed by bright excitons along the molecular chain.
\begin{figure}[h!]
\centering
\includegraphics[width=\linewidth]{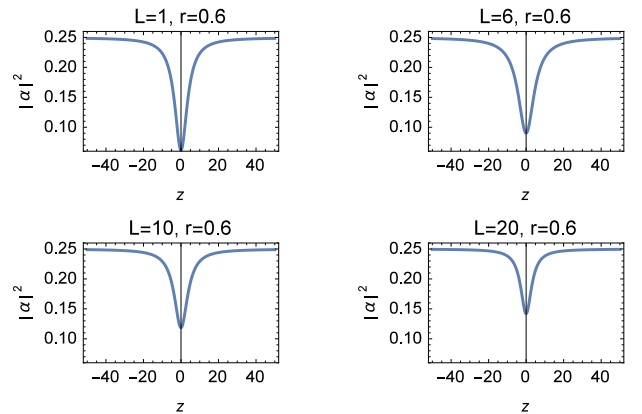}
  \caption{ The squared modulus of the dark soliton solution versus $z$ for four different values of the interaction range, $L$, i.e. $L=1$, $6$, $10$, $20$: $r=0.6$.
 }
  \label{fig:Fig5}
\end{figure}

\begin{figure}[h!]
\centering
\includegraphics[width=\linewidth]{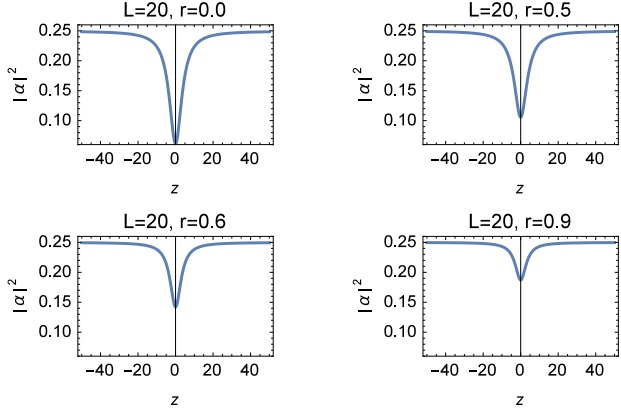}
  \caption{ The squared modulus of the dark soliton solution versus $z$ for four different values of the interaction strength, $r$, i.e. $r=0.0$, $0.5$, $0.6$, $0.9$: $L=20$.
 }
  \label{fig:Fig6}
\end{figure}
Turning to the dark soliton solution (\ref{Nsol2}), in fig. \ref{fig:Fig5} we sketched the squared modulus of the associate occupation probability versus $z$ for the same values of $L$ and $r$ considered in the bright-soliton case. Fig. \ref{fig:Fig6} shows plots of the same function but for $L = 20$, and considering four different values of $r$ i.e. $r=0.0$, $0.5$, $0.6$, $0.9$. Other characteritic parameters of the model assume the same numerical values we picked for the bright-soliton case. One sees that as the intermolecular interaction range $L$ is gradually increased from the first nearest neighbours (fig. \ref{fig:Fig5}), amplitude of the dark soliton undergoes a huge decrease. As the strength of intermolecular interactions $r$ is gradually increased for $L=20$ (fig. \ref{fig:Fig6}), the dark soliton amplitude decreases slightly in the region $0\leq r \leq0.5$ (i.e. for weak intermolecular coupling), and undergoes a significantly huge decrease in the region $0.5< r\leq1$ (i.e. for strong intermolecular coupling). These behaviours too are quite revealing of a significant quantitative role expected from long-range interactions, in the energy transfer mediated by dark-soliton profile excitons. 
\begin{figure}[h!]
\centering
\includegraphics[width=3.3in]{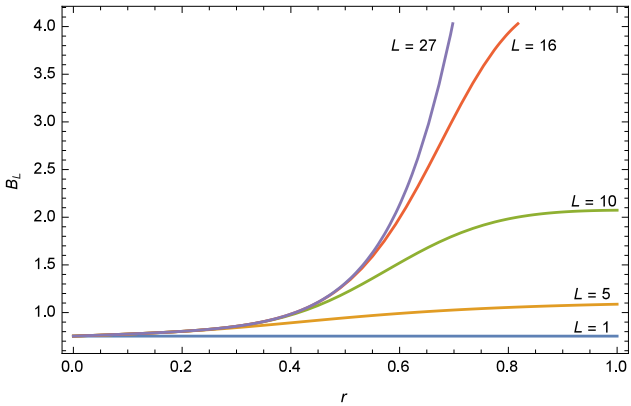}
  \caption{The soliton parameter, $B_L$, plotted in the long-wavelength regime (q $\approx$ 0.05) versus
  the long-range interaction strength $r$, for five different values of the interaction range, $L$, i.e. $L = 1$, $5$, $10$, $16$ and $27$,
  in increasing order starting from the horizontal curve. Other parameter values are; $J_0 = 1$, $\omega_0 = 0.9$, $\omega = 1.2$, $\Omega_0 = 0.1$, $a = 1$, $\chi = 0.3$ and $v = 0.2$.
 }
  \label{fig:Fig7}
\end{figure}

Before concluding, it is instructive to note that in both the bright and dark soliton solutions obtained above, there was a parameter denoted $B_L$ which clearly carries the dependence of the soliton characteristic amplitude on the long-range parameters $L$ and $r$. For the expression of this parameter suggests a complex dependence on the two characteristic long-range parameters, it is useful to look at its behaviour as a function of these two parameters. Thus, fig. \ref{fig:Fig7} shows the variation of $B_L$ in the long-wavelength regime, versus the long-range interaction strength $r$ for five different values of the interaction range namely $L = 1$, $5$, $10$, $16$, $27$. The horizontal curve corresponds to the first nearest neighbour regime, where from fig. \ref{fig:Fig7} it seen that in this case $B_L < 1$. As
the range and strength of intermolecular interactions are gradually increased,
the parameter $B_L$ increases from its short-range value and becomes greater
than 1 ($B_L > 1$). It saturates beyond a finite value of the intermolecular interaction range $L$.
The observed switch from $B_L < 1$ to $B_L > 1$, triggered by the sensitivity to long-range
interactions, corresponds to a switch from the red frequency shift to a blue
frequency shift of the central carrier wave frequency of the
exciton-polariton \cite{prim2002}. It turns out that taking into account the light-induced nonlinear polarization of the medium leads to the generation of elliptic-type single-exciton-photon solitary wave structures, in which long-range intermolecular interactions strongly control amplitude and phase modulations of the solitary wave solutions. So to say, in the long-wavelength regime long-range intermolecular interactions could be an important part of the underlying mechanism responsible for the formation of chirped exciton-polariton solitary waves in one-dimensional molecular crystals.

\section{Concluding remarks}
In this work we have investigated the effects of long but finite range dispersive intermolecular interactions, on soliton bound states consisting of excitons and polaritons in one dimensional molecular crystals with strong lattice-field interactions. A modified nonlinear Schr\"odinger equation was obtained whose nonlinear and dispersive coefficients were significantly enhanced by long-range intermolecular interactions. Elliptic-type bright and dark solitary-wave solutions to this nonlinear equation were obtained analytically. These solutions are periodic trains of bright or dark solitons that maintain a steady shape profile, irrespective of the value of the intermolecular coupling strength. Long-range intermolecular interactions were found to induce a sizeable increase in the bright solitary wave amplitude, and a decrease in the dark solitary wave amplitude, with a quasi-doubling and quasi-halving of the respective amplitude in the strong long-range interaction regime. Nevertheless these changes in the solitary wave amplitudes as the intermolecular interaction range is increased, were found to be more subtle in the limit of weak intermolecular coupling strength. We also found that long-range interactions induce an increase in the soliton parameter $B_L$, which corresponds to a switch from the red frequency shift to a blue frequency shift of the central carrier wave frequency of the
exciton-polariton boundstate. These results can be understood in that in the long wavelength regime, long-range intermolecular interactions could contribute as an underlying mechanism responsible for the formation of amplitude and phase modulated exciton-polariton solitary waves in molecular crystals.

\section*{Authors contribution statement}
ENNA performed calculations, AMD proposed the theme and wrote the manuscript. Both authors agreed on curves which were plotted by AMD.
\section*{Conflict of interest}
The authors declare that they have not conflict of interest.
%

\end{document}